\documentclass[11pt]{article}

\usepackage[margin=2.5cm]{geometry}
\usepackage{amsmath,amssymb,amsfonts,mathtools}
\usepackage{bm}
\usepackage{physics}
\usepackage{graphicx}
\usepackage{hyperref}
\usepackage{comment}
\usepackage{authblk}
\usepackage{booktabs}
\usepackage{array}
\usepackage{tabularx}
\usepackage{amsmath}
\usepackage{booktabs}
\usepackage{comment}
\usepackage{array}
\usepackage{dcolumn}
\usepackage{bm}
\usepackage{hyperref}

\title{The Reweighting Principle in Statistical Mechanics}

\author[1,2]{Salvatore Romano}

\affil[1]{University of Vienna, Faculty of Physics, Boltzmanngasse 5, 1090 Vienna, Austria}
\affil[2]{University of Vienna, Vienna Doctoral School in Physics (VDSP) Boltzmanngasse 5, 1090, Vienna, Austria}
\date{}

\begin{document}

\maketitle

\begin{abstract}
Reweighting of probability measures provides a unifying perspective on ensemble transformations in statistical mechanics. We distinguish two complementary classes of reweighting: soft constraints, which redistribute probability while preserving the support of the reference measure, and hard constraints, which impose support restrictions through conditioning. We show that exponential tilting and conditioning on an exact observable value arise as the minimum relative entropy updates associated with soft expectation constraints and hard exact-value constraints, respectively. Their relative entropies naturally inherit complementary thermodynamic structures: exponential tilting gives rise to the Legendre structure of the canonical ensemble and reduces, for a uniform reference measure, to Gibbs entropy, whereas conditioning reduces to Boltzmann entropy through the surprisal of the constrained macrostate. By introducing an enlarged probability space in which observables are treated as explicit random variables, we further show that canonical and microcanonical ensembles arise as marginal and conditional distributions of a common joint reference measure. In the thermodynamic limit, large-deviation concentration makes soft and hard constraints macroscopically equivalent, providing a probabilistic interpretation of canonical--microcanonical ensemble equivalence. Finally, we outline how the same information-theoretic framework naturally extends to path space, suggesting a unified probabilistic description of equilibrium statistical mechanics and conditioned stochastic dynamics.
\end{abstract}

\section{Introduction}

Reweighting of probability measures is a fundamental operation underlying a wide range of constructions in statistical mechanics, stochastic processes, and inference \cite{FrenkelSmit2002}. The idea is remarkably simple. Given a reference prior, a new ensemble can be obtained by modifying the relative likelihood of configurations through a non-negative weight function and a normalization factor. This elementary operation appears in many forms. It underlies free-energy calculation methods \cite{Zwanzig1954,Bennett1976}, importance sampling and enhanced-sampling approaches such as umbrella sampling \cite{Torrie1977}, histogram and multistate reweighting techniques \cite{Ferrenberg1989,Kumar1992,Shirts2008,WuNoe2014}, and modern generative modeling approaches to equilibrium sampling \cite{Rezende2015,Noe2019, park_deep_2025}.

Despite its ubiquity, reweighting is often introduced with different interpretations in different fields. In probability theory, it appears as a change of measure \cite{Kallenberg2002, Billingsley1995, Bradley01051989}. From an information-theoretic perspective, it can be viewed as a generalized Bayesian update, in which prior probabilities are modified by newly imposed information \cite{Jaynes1957,KapurKesavan1992}. In statistical mechanics, exponential weights define thermodynamic ensembles and couple observables to conjugate fields\cite{VanCampenhoutCover1981,Csiszar1984}.  On path space, analogous
changes of measure underlie conditioned processes, entropy production,
fluctuation relations, and rare-event ensembles
\cite{YangQian2020,ChetriteTouchette2015,Touchette2018}.
Related constructions appear as maximum entropy or minimum relative entropy updates
\cite{Kullback1951,Jaynes1957,VanCampenhoutCover1981,Csiszar1984,
KapurKesavan1992}, while large deviation theory provides the asymptotic
framework connecting tilted ensembles, entropy, and thermodynamic limits \cite{Vasicek1980,VanCampenhoutCover1981,Csiszar1984,
DemboZeitouni1998,Ellis1985,Touchette2009}. These viewpoints emphasize different aspects of the same operation, but they are rarely presented within a common framework.

The aim of this work is to organize these structures around the
reweighting of a reference distribution or measure. A nonnegative
weight separates naturally into two components: its support determines which
configurations remain admissible, while its positive part redistributes
probability within that support. Reweighting therefore provides the general
algebraic form of an ensemble transformation, encompassing both hard
restrictions and ordinary changes of measure.

Minimum relative entropy provides the corresponding prescriptive framework:
imposed information selects the least-biased ensemble and thereby determines
its reweighting factor. Conversely, a prescribed weight defines an information
potential and an associated variational representation. This parallel
description places soft constraints, implemented by exponential tilting, and
hard constraints, implemented by conditioning, within a common framework and
clarifies their respective connections with Gibbs and Boltzmann entropy.

We then introduce an enlarged-space representation in which microscopic
configurations and observable values are treated jointly. This construction
allows tilted and conditioned ensembles to be compared as different
probabilistic operations on the same joint measure. Finally, we examine their
relation in the thermodynamic limit, where large-deviation concentration
provides the link between soft and hard constraints. Viewed in this way,
reweighting connects the physical information defining an ensemble to the
support restrictions and probability transformations through which that
information is imposed.

\section{The Reweighting Principle}
\label{sec:reweighting}
Given a reference ensemble \(\rho_0(x)\), a new probability density is obtained by modifying the relative likelihood of configurations through a non-negative weight \(w(x)\):
\begin{equation}
\rho_w(x)
=
\frac{w(x)\rho_0(x)}{\langle w\rangle_0},
\qquad 
\langle w\rangle_0
=
\int dx\,w(x)\rho_0(x).
\label{eq:general_reweighting}
\end{equation}
The weight changes the relative likelihood of configurations, while the
normalization, assumed such that $0<\langle w\rangle_0<\infty$, ensures that the resulting measure remains a probability measure. Although $\rho_0$ is taken here to be normalized, the reweighting construction applies to an underlying measure that is not necessarily normalized, provided that the weighted integral in the denominator is finite. We retain the probability-density notation for simplicity of the discussion and return to the normalization requirement when introducing relative entropy in the next section.

The power of Eq.~\eqref{eq:general_reweighting} is that all information about the ensemble transformation is encoded in a single object, the weight \(w(x)\): different choices of \(w\) generate different ways of incorporating information into a reference ensemble. In probability theory terms, Eq.~\eqref{eq:general_reweighting} corresponds to a change of probability measure with Radon–Nikodym derivative proportional to $w$ but it can be interpreted as a generalization of Bayesian updating and inference \cite{Kallenberg2002, Billingsley1995, YangQian2020}. A general nonnegative weight combines a hard restriction of the accessible
configurations with a soft redistribution of probability within the retained support. To make this separation explicit, let us decompose the weight function as
\begin{equation}
w(x)
=
1_{E_w}(x) \widetilde w(x),
\label{eq:weight_decomposition}
\end{equation}
where $1_{E_w}$ is the indicator of the support of the weight $E_w=\{x:w(x)>0\}$ and $\widetilde w(x)=w(x)>0$ on $E_w$. The indicator
function restricts the ensemble to the admissible event $E_w$,
whereas $\widetilde w$ modifies the relative probabilities of configurations
that remain accessible. The distribution obtained by reweighting the reference ensemble with the indicator function is
\begin{equation}
\rho_{0,E_w}(x)
=
\frac{1_{E_w}(x)\rho_0(x)}
{P_0(E_w)}=\rho_0(x\mid E_w).
\label{eq:conditional_reference_weight}
\end{equation}
Its probabilistic interpretation is Bayesian conditioning: the reference ensemble plays the role of the prior, while the indicator encodes the event that has been observed. Their product gives the unnormalized updated distribution, and normalization yields the conditional ensemble. Given this, the full reweighted distribution in Eq.~\ref{eq:general_reweighting} can be seen as
\begin{equation}
\rho_w(x)
=
\frac{\widetilde w(x)\rho_{0,E_w}(x)}
{\langle\widetilde w\rangle_{0,E_w}},
\label{eq:hard_soft_reweighting}
\end{equation}
where 
\begin{equation}
\langle w\rangle_0
=
P_0(E_w)
\langle\widetilde w\rangle_{0,E_w}.
\end{equation}
Thus, every reweighting can be interpreted as conditioning on the
support of the weight (hard constraint) followed by redistributing probabilities within that support (soft constraint). Zeros determine which configurations are excluded, while the positive part determines the relative likelihoods of the remaining possible configurations.

Among the many possible reweightings, two constructions play a distinguished
role in statistical mechanics and probability theory. 

\paragraph{Soft constraint: exponential tilting.}
A smooth reweighting of the reference ensemble can be obtained by introducing an
exponential bias of an observable \(A(x)\). 
\begin{equation}
w_\lambda(x)=e^{-\lambda A(x)}.
\label{eq:exp_weight}
\end{equation}
This gives the tilted ensemble
\begin{equation}
\rho_{\lambda}(x)
=
\frac{e^{-\lambda A(x)}\rho_0(x)}{Z_0(\lambda)},
\label{eq:exp_rew}
\end{equation}
where the normalization 
\begin{equation}\label{eq:partition_function}
Z_0(\lambda)
=
\int dx\,\rho_0(x)e^{-\lambda A(x)}.
\end{equation}
This construction plays a central role in statistical mechanics, where it
underlies the definition of canonical ensembles through the duality between the
observable \(A\) and its conjugate field \(\lambda\)
\cite{VanCampenhoutCover1981,Csiszar1984,DemboZeitouni1998,Touchette2009}. 
Exponential tilting naturally generates transformations between
thermodynamic ensembles of both intensive and extensive variables (Appendix~\ref{app:switch}).
 
Here and throughout, the thermodynamic interpretation is obtained by choosing the energy as observable \(A(x)=H(x)\), the inverse temperature as conjugate variable \(\lambda=\beta\), and the phase-space measure as the underlying reference measure. In this case, Eq.~\ref{eq:exp_rew} reduces to the usual canonical ensemble and Eq.~\ref{eq:partition_function} to the partition function.

\paragraph{Hard constraint: exact conditioning.}
Conditioning on an event of nonzero reference probability is represented
directly by an indicator weight. Prescribing instead an exact value
$\bar A$ of a continuous observable $A(x)$ corresponds formally to
conditioning on the level set $A(x)=\bar A$, represented by the
Dirac-delta weight
\begin{equation}
w_{\bar A}(x)=\delta(\bar A-A(x)).
\label{eq:delta_weight}
\end{equation}
The reweighted density is then
\begin{equation}
\rho_{\bar A}(x)
=
\frac{\delta(\bar A-A(x))}{\Omega_0(\bar A)}\,\rho_0(x),
\label{eq:conditioning}
\end{equation}
where 
\begin{equation}\label{eq:density_of_st}
\Omega_0(\bar A)
=
\int dx\,\rho_0(x)\delta(\bar A-A(x)),
\end{equation}
is the total mass of the measure induced by the reference ensemble on the
level set $(A(x)=\bar A)$. Because this exact level set generally has zero
probability under the original reference ensemble, the resulting conditional
measure is singular with respect to it.  The Dirac-delta expression can be
understood as the sharp-shell limit of conditioning on a finite resolution
macrostate \cite{Touchette2015} (see
Appendix~\ref{app:min_rel}). 

Eq.~\ref{eq:conditioning} has the thermodynamic structure of the microcanonical ensemble with Eq.~\ref{eq:density_of_st} as density of states, when the reference density is constant with respect to the phase-space measure \cite{Balian2007}.
The reweighted ensemble is therefore the posterior distribution conditioned on the event \(A(x)=\bar A\),
\begin{equation}
\rho_{\bar A}(x)
=
\rho_0\bigl(x \mid A(x)=\bar A\bigr).
\end{equation}
Thus, hard reweighting with a delta is simply the probabilistic operation
of incorporating exact information into a reference ensemble.
The Bayesian form of conditioning suggests that the relevant object is a joint
distribution over configurations and observables \cite{Touchette2015}. This observation motivates the
enlargement of the space \((x,A)\) introduced in Sec.~\ref{sec:enlarged}.

\medskip

\begin{figure*}[h]
    \centering
    \includegraphics[width=0.9\linewidth]{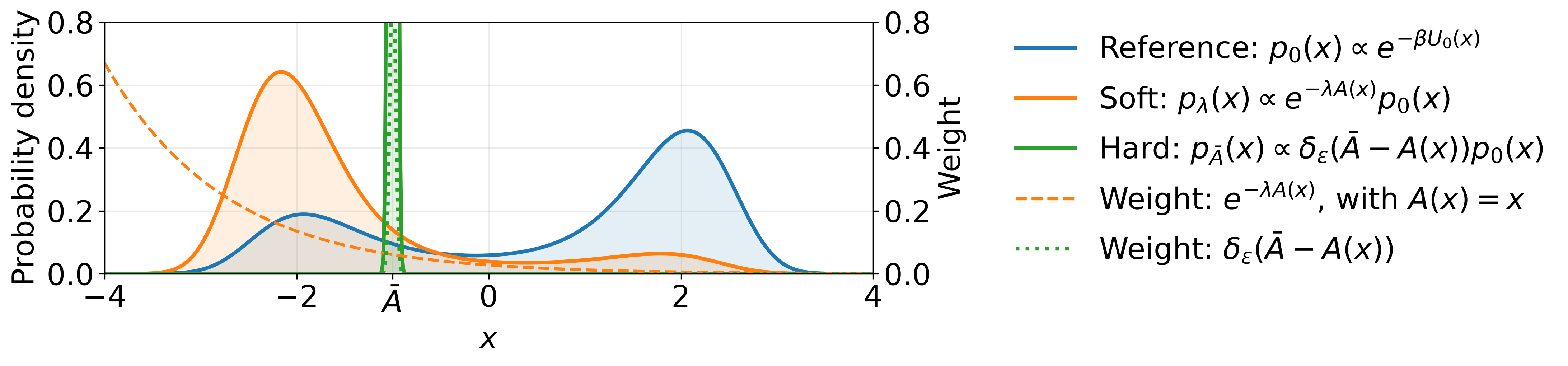}
    \caption{Schematic illustration of soft and hard reweighting. Exponential tilting (orange)
modifies the reference ensemble (blue) smoothly through an exponential weight, while
conditioning (green) selects a finite resolution macrostate associated with a prescribed
value of the observable.}
    \label{fig:soft_hard}
\end{figure*}

The contrast between these two elementary ensemble transformations is illustrated schematically in Fig.~\ref{fig:soft_hard}.  Exponential tilting reshapes the reference ensemble by changing the relative likelihood of configurations, whereas conditioning restricts the ensemble to configurations compatible with a prescribed value of the observable. Having characterized these reweightings, we now turn to their informational interpretation in terms of relative entropy and to the variational problem that select them.

\section{Reweighting and Minimum Relative Entropy}
\label{sec:min_relative_entropy}
The reweighting formula in Eq.~\eqref{eq:general_reweighting} is operational rather than
prescriptive. It specifies how an ensemble changes once the weight is given, but
does not determine which weight should be chosen when new information is available about the system. 
The principle of minimum relative entropy addresses this selection problem: among all distributions compatible with the prescribed information $I$, $\mathcal{R}_I=\{ \rho\ \mid  \rho \text{ satisfies I} \}$, it selects the one closest to the reference ensemble in Kullback--Leibler divergence (or relative entropy) through \cite{Kullback1951,  Jaynes1957, KapurKesavan1992}
\begin{equation}
\rho_{I}
=
\operatorname*{arg \ min}_{\rho\in \mathcal{R}_I}
D(\rho \mid\rho_0),
\qquad
D(\rho\mid\rho_0)
=
\int dx\rho(x)\log\frac{\rho(x)}{\rho_0(x)} 
\label{eq:min_relative_entropy}
\end{equation}
Although $\rho_0$ is taken here to be a normalized probability density, the
same variational construction can be formulated relative to a reference measure that need not be normalizable \cite{Perlaza2024}. The relative entropy with respect to a measure $m_0(dx)=\mu_0(x)dx$ is defined using its density $\mu_0(x)$. When $m_0$ has finite total mass, it differs from the Kullback--Leibler divergence with respect to its normalized version only by an additive constant. For an infinite reference measure, no normalized prior exists, but the variational problem remains meaningful if the relative entropy functional and the associated normalization are finite on a non-empty admissible class of densities.

Whenever the minimizing distribution is absolutely continuous with respect to the reference ensemble, i.e. $\rho_I(x)=0$ if $\rho_0(x)=0$, it can be expressed in the reweighted form \eqref{eq:general_reweighting}. Indeed, the corresponding weight can be identified as
\begin{equation}
w_I(x)\propto\frac{\rho_I(x)}{\rho_0(x)},
\end{equation}
where the proportionality constant is irrelevant because it cancels upon
normalization. 


To clarify how a weight is related to information imposed through constraints,
we recall the Gibbs--Donsker--Varadhan variational representation, which
establishes a fundamental connection between positive reweighting and relative
entropy minimization \cite{DonskerVaradhan1975,DupuisEllis1997,Hartmann2017,park_deep_2025}. Given a strictly positive weight $w(x)$, let us define the effective information potential
\begin{equation}
U_w(x)=-\log w(x).
\end{equation}
For any normalized probability density \(\rho\), the relative entropy with
respect to the reweighted ensemble can be written as
\begin{align}
D(\rho\mid\rho_w)
=
D(\rho\mid\rho_0)
+
\langle U_w\rangle_\rho
+
\log \langle w \rangle_0 .
\label{eq:general_reweighting_variation}
\end{align}
Since \(D(\rho \mid \rho_w)\geq0\), with equality only when
\(\rho=\rho_w\), it follows that
\begin{equation}
\rho_w
=
\operatorname*{arg\,min}_{\rho}
\left[
D(\rho\mid\rho_0)
+
\langle U_w\rangle_\rho
\right],
\label{eq:general_reweighting_min}
\end{equation}
and
\begin{equation}
-\log \langle w \rangle_0
=
\min_{\rho}
\left[
D(\rho\mid\rho_0)
+
\langle U_w\rangle_\rho
\right].
\end{equation}
Thus, a prescribed positive reweighting can be understood as a balance between
the relative entropy cost of departing from the reference ensemble and the
preference encoded by the logarithmic weight.

More generally, introducing a multiplier \(\lambda\) before the
average potential term gives
\begin{equation}
\frac{w(x)^\lambda\rho_0(x)}
{\langle w^\lambda\rangle_0}=\operatorname*{arg\,min}_{\rho}
\left[
D(\rho\mid\rho_0)
+
\lambda\langle U_w\rangle_\rho
\right].
\end{equation}
This is the Lagrangian form of a minimum relative entropy problem with a
prescribed value of \(\langle U_w\rangle_\rho\). The original reweighted
ensemble \(\rho_w\) is recovered for \(\lambda=1\), corresponding to the
constraint value \(\langle U_w\rangle_\rho=\langle U_w\rangle_{\rho_w}\).

The strictly positive case describes a support-preserving change of measure. To extend the variational representation to nonnegative weights, we
recall the decomposition introduced in
Eq.~\eqref{eq:weight_decomposition} and the corresponding ensemble in
Eq.~\eqref{eq:hard_soft_reweighting}, obtained
by first conditioning the reference ensemble on the support $E_w$ and then applying the
positive weight $(\widetilde w)$ within the restricted support.

This algebraic separation has a direct information-theoretic counterpart in terms of relative entropy
\begin{equation}
D(\rho_w\mid\rho_0)
=
-\log P_0(E_w)
+
D(\rho_w\mid\rho_{0,E_w}).
\label{eq:hard_soft_information_decomposition}
\end{equation}
where $\rho_{0,E_w}=\rho_0(\cdot\mid E_w)$. The first term is the
surprisal associated with restricting the reference ensemble to the admissible
event, whereas the second measures the additional redistribution of
probability within that event. Thus, the information introduced by a
nonnegative weight separates into a hard contribution determined by its zeros
and a soft contribution determined by its positive part.

The Gibbs--Donsker--Varadhan representation can then be applied to
$\widetilde w$ relative to the conditioned reference ensemble
$\rho_{0,E_w}$. Equivalently, the extended information potential
$U_w=-\log w$ is infinite outside $E_w$, thereby enforcing the hard support
constraint, and finite within $E_w$, where it generates the remaining soft
bias. For an indicator weight, $\widetilde w=1$, the second contribution in
Eq.~\eqref{eq:hard_soft_information_decomposition} vanishes and the update
reduces to pure conditioning.

There are therefore two complementary variational directions. In the
inference direction, information is specified through a constraint set, and
the principle of minimum relative entropy determines the corresponding
ensemble and its reweighting weight. Conversely, when a weight is prescribed,
the Gibbs--Donsker--Varadhan principle provides a relative entropy
characterization of the resulting ensemble, while the support decomposition
distinguishes the information associated with excluding configurations from
that associated with changing their relative likelihoods.

We now retrieve the two forms of reweighting introduced in the previous section starting with physically prescribed information and use minimum relative entropy to determine the corresponding weight. Expectation-value constraints generate exponential tilting, whereas
event or support constraints generate conditioning.

\paragraph{Exponential tilting and Gibbs entropy}
A soft constraint fixes the expectation value of an observable,
\begin{equation}
    \langle A\rangle_\rho=\bar A .
\end{equation}
The least biased ensemble compatible with this information is obtained from
 minimizing the relative entropy over all distributions satisfying the constraint. The solution of this constrained variational problem, derived explicitly in Appendix \ref{app:min_rel}, is the exponential family in Eq.~\ref{eq:exp_rew}.
In reweighting language, the minimum relative entropy update
associated with an average constraint is generated by the observable exponential weight in Eq.~\ref{eq:exp_weight}.

The relative entropy of the
tilted ensemble is
\begin{equation}
D(\rho_\lambda\mid\rho_0)
=
-\lambda\langle A\rangle_\lambda
-
\log Z_0(\lambda)
\label{eq:kl_tilt}
\end{equation}
which displays the Legendre structure of equilibrium statistical
mechanics with the normalization $Z_0(\lambda)$ playing the role of a partition function, while its logarithm of the corresponding free energy-like potential.

The thermodynamic interpretation becomes transparent when the reference measure is constant. For a finite or coarse-grained state space, the corresponding normalized reference ensemble is the uniform distribution, and the relative entropy becomes
\begin{equation}
D(\rho_\lambda\mid\rho_{\mathrm{unif}})
=
-S_G(\rho_\lambda)+\mathrm{const.},
\end{equation}
where
\begin{equation}
S_G(\rho_\lambda)
=
-\int dx\,\rho_\lambda(x)\log\rho_\lambda(x)
\end{equation}
is the Gibbs entropy. Throughout, Boltzmann’s constant is set to unity. More generally, for a constant reference measure that
need not be normalizable, the corresponding relative entropy functional is
defined directly with respect to that measure and reduces again to $-S_G$ up to an
additive constant whenever the measure has finite mass. Hence, Gibbs entropy
maximization is recovered as the constant reference measure case of minimum
relative entropy.

\paragraph{Conditioning and Boltzmann entropy}
A hard constraint fixes the exact value of an observable rather than an expectation value. Formally, the support of the distribution is restricted to configurations satisfying
\begin{equation}
A(x)=\bar A.
\end{equation}
As shown in Appendix~\ref{app:min_rel}, the solution of the corresponding minimum relative entropy problem is the conditional distribution in Eq.~\ref{eq:conditioning}, which, in reweighting language, is the update generated by the generalized weight in Eq.~\ref{eq:delta_weight}.

To assign a finite relative entropy to this conditioning operation, we
regularize the exact constraint by introducing the finite resolution shell of width \(\varepsilon\), exactly as in the classical microcanonical ensemble
\begin{equation}
    \bar A_{\varepsilon}=\{x:\bar A\le A(x)\le\bar A+\varepsilon\}.
\end{equation}
Now, on the support of the conditioned ensemble, the relative entropy between the conditioned and reference ensembles reduces to the surprisal of the constrained macrostate:
\begin{equation}
D(\rho_{\bar A_{\varepsilon}}\mid\rho_0)
=
-\log P_0(\bar A_{\varepsilon})
\label{eq:kl_conditioning}
\end{equation}
For a sufficiently thin
shell, the probability assigned to the shell
\(\bar A_\varepsilon\) is proportional to the reference density of states, 
\begin{equation}
P_0(\bar A_\varepsilon)
=\Omega_0(\bar A)\,\varepsilon+o(\varepsilon).
\end{equation}
Consequently,
\begin{equation}\label{eq:boltz_rel_ent}
D(\rho_{\bar A_\varepsilon}|\rho_0)
=-\log\varepsilon
-\log\Omega_0(\bar A)
+\mathrm{o(1)}
\end{equation}
The divergent term is the universal information cost of imposing an increasingly sharp constraint, whereas the second is the finite, macrostate-dependent contribution. 
In the limit $\varepsilon\to0$, for a uniform distribution over a finite or coarse-grained state space, the finite part of the relative entropy becomes
\begin{equation}
D_{\mathrm {fin}}(\rho_{\bar A} \mid\rho_{\mathrm{unif}})
=
-S_B(\bar A)+ \mathrm{const.},
\end{equation}
where we retained only the finite, thermodynamical relevant contribution (second term in Eq.~\ref{eq:boltz_rel_ent}) and we recognize the Boltzmann entropy
\begin{equation}
S_B(\bar A)=\log\Omega(\bar A).
\end{equation}
This is the energy-shell realization of the Boltzmannian definition of entropy
as the logarithm of the phase-space volume associated with a macrostate
\cite{Lebowitz1999}. In this interpretation, Boltzmann entropy characterizes
the multiplicity of microscopic configurations compatible with prescribed
macroscopic information.
As in the Gibbs-entropy case, the same relation can
be formulated directly with respect to a constant reference measure that need
not be normalizable. Its arbitrary normalization contributes only an additive
constant independent of the macrostate. 

\medskip

The two constructions therefore place Gibbs and Boltzmann entropy within the
common information-theoretic framework of relative entropy, distinguished
between expectation and support constraints. This unifies the canonical and microcanonical constructions that admit perfectly parallel structure summarized in Table~\ref{tab:soft_hard}.

\begin{table}[h]
\centering
\small
\renewcommand{\arraystretch}{1.25}

\begin{tabularx}{\linewidth}{
@{}
>{\raggedright\arraybackslash}p{0.19\linewidth}
>{\centering\arraybackslash}X
>{\centering\arraybackslash}X
@{}
}
\toprule
&
\textbf{Soft constraint}
&
\textbf{Hard constraint}
\\
\midrule

\multicolumn{3}{@{}l}{\textbf{Reweighting structure}}
\\[2pt]

Operation
&
Exponential tilting
&
Conditioning
\\[3pt]

Weight
&
$\displaystyle
w_\lambda(x)=e^{-\lambda A(x)}$
&
$\displaystyle
w_{\bar A}(x)=\delta\!\left(\bar A-A(x)\right)$
\\[6pt]

Normalization
&
$\displaystyle
Z_0(\lambda)
=
\left\langle e^{-\lambda A}\right\rangle_0$
&
$\displaystyle
\Omega_0(\bar A)
=
\left\langle
\delta\!\left(\bar A-A(x)\right)
\right\rangle_0$
\\[8pt]

Reweighted ensemble
&
$\displaystyle
\rho_\lambda(x)
=
\frac{e^{-\lambda A(x)}}{Z_0(\lambda)}
\rho_0(x)$
&
$\displaystyle
\rho_{\bar A}(x)
=
\frac{\delta\!\left(\bar A-A(x)\right)}
{\Omega_0(\bar A)}
\rho_0(x)$
\\[8pt]

Effect
&
Smooth redistribution of probability
&
Singular restriction to a level set
\\

\midrule
\multicolumn{3}{@{}l}{\textbf{Thermodynamic structure}}
\\[2pt]

Information imposed
&
$\langle A\rangle_\rho=\bar A$
&
$A(x)=\bar A$
\\
Relative entropy
&
$\displaystyle
D(\rho_\lambda\mid\rho_0)
=
-\lambda\langle A\rangle_\lambda
-\log Z_0(\lambda)$
&
$\displaystyle
D(\rho_{\bar A_\varepsilon}\mid\rho_0)
=
-\log P_0(\bar A_\varepsilon)$
\\[8pt]

Information structure
&
Legendre duality
&
Surprisal of the event
\\[3pt]

Constant-reference limit
&
$\displaystyle
D(\rho_\lambda\mid\rho_{\mathrm{unif}})
=
-S_G(\rho_\lambda)+\mathrm{const.}$
&
$\displaystyle
D_{\mathrm{fin}}(\rho_{\bar A}\mid
\rho_{\mathrm{unif}})
=
-S_B(\bar A)+\mathrm{const.}$
\\[8pt]

Thermodynamic realization
&
Canonical ensemble
&
Microcanonical ensemble
\\
\bottomrule
\end{tabularx}

\caption{Parallel structure of soft and hard constraint as reweighting operations (upper block) and minimum relative entropy updates (lower block). Conditioning is represented formally by a Dirac-delta weight. Because the exact conditional measure is singular with respect to the original phase-space measure, its relative entropy is evaluated using a finite resolution shell. In the thermodynamic interpretation, only the finite, macrostate-dependent part is retained after subtracting the universal resolution-dependent contribution.}
\label{tab:soft_hard}
\end{table}


\section{Enlarged Space Representation of Thermodynamics}
\label{sec:enlarged}
The previous sections showed that exponential tilting naturally generates the
canonical ensemble, while conditioning provides the microcanonical analogue.
The Bayesian interpretation of conditioning suggests making explicit the joint
distribution of the microscopic configuration and the observable on which one
conditions \cite{Touchette2015}. What one obtains is a
natural probability space from which canonical and microcanonical ensembles
emerge as different probabilistic projection.

To make this construction explicit, consider the enlarged space \((x,A)\), where the observable is promoted as an explicit random variable. To lighten the notation, from now on, the symbol ``$A$'' is used both for the observable $A(x)$ and for its value as a coordinate of the enlarged space; the distinction will always be clear from the argument. The conditioning factor \(\delta(A-A(x))\) is not solely a formal restriction but it can be used to define the joint probability law 
\begin{equation}
\label{prior}
\rho_0(x,A)
=
\delta(A-A(x)) \rho_0(x).
\end{equation}
This measure encodes the prior knowledge of the system and the deterministic relation induced by the map \(x\mapsto A(x)\), without introducing any additional statistical bias. 

Exponential reweighting in the observable sector gives the joint distribution:
\begin{equation}\begin{split}\label{eq:joint}
\rho_\lambda(x,A)
&=
\frac{e^{- \lambda A}}{\langle e^{- \lambda A }\rangle_0 }\,\rho_0(x,A)
=
\frac{\delta(A-A(x))\,e^{- \lambda A} \rho_0(x)}{Z_0(\lambda)},
\end{split}
\end{equation}
with partition function
\begin{equation}
Z_0(\lambda)
=
\int dx\,dA\, e^{- \lambda A}\delta(A-A(x))\rho_0(x)
\end{equation}
The probabilistic structure of the enlarged space becomes transparent by
examining the two marginals and the two conditional distributions associated
with the joint \(\rho_\lambda(x,A)\). 

\paragraph{Configuration marginal.}
Marginalizing over the observable gives
\begin{equation}
\begin{split}
\rho_\lambda(x)=
\frac{e^{-\lambda A(x)}\rho_0(x)}
{Z_0(\lambda)}.
\end{split}
\label{eq:configuration_marginal}
\end{equation}
Thus, exponential reweighting in the observable coordinate induces the
corresponding tilted ensemble in configuration space.

\paragraph{Observable marginal.}
Marginalizing instead over the configurations gives
\begin{equation}
\rho_\lambda(A)
=
\frac{e^{-\lambda A}\Omega_0(A)}
{Z_0(\lambda)}.
\label{eq:observable_marginal}
\end{equation}
This distribution describes the competition between the reference weight of
a given observable value and the exponential bias.

\paragraph{Configuration conditional.}
Conditioning the joint distribution on a fixed value of the observable gives
\begin{equation}
\begin{split}
\rho_\lambda(x\mid A)
=
\frac{\delta(A-A(x))\rho_0(x)}
{\Omega_0(A)}
=
\rho_0(x\mid A).
\end{split}
\label{eq:configuration_conditional}
\end{equation}
The exponential factor cancels because it is constant on the level set $A(x)=A$. Exponential reweighting therefore changes the relative
probabilities of different observable values, but leaves the conditional
distribution within each level set unchanged.

\paragraph{Observable conditional.}
Conversely, conditioning on a microscopic configuration reduces to the deterministic relation between the configuration and its observable
\begin{equation}
\rho_\lambda(A\mid x)
=
\delta(A-A(x)).
\label{eq:observable_conditional}
\end{equation}

The enlarged space representation therefore provides a unified probabilistic
description of tilting and conditioning: exponential reweighting modifies the
distribution among observable values, whereas conditioning selects a
particular level set independently of the tilt.

In the thermodynamic specialization, for a flat phase-space reference measure, the configuration marginal Eq.~\eqref{eq:configuration_marginal} becomes the canonical ensemble, whereas
the configuration conditional Eq.~\eqref{eq:configuration_conditional}
becomes the microcanonical ensemble at fixed energy.

\section{Thermodynamic Limit and Emergence of Conditioning}
\label{sec:therm_lim}
The classical equivalence between canonical and microcanonical ensembles can
be interpreted, within the present framework, as a macroscopic equivalence
between soft and hard reweighting. The key point is that, in the thermodynamic
limit, the marginal distribution of the observable concentrates, causing the
exact mixture of conditional ensembles to become dominated by a single
conditional component.

Let \(N\) denote the number of particles and \(a(x)=A(x)/N\) the intensive
macrostate. The joint distribution in
Eq.~\eqref{eq:joint} can then be factorized as
\begin{equation}
\rho_{N,\lambda}(x,a)
=
\rho_{N,\lambda}(a)\,\rho_N(x\mid a).
\end{equation}
Consequently, for every finite \(N\), the tilted configuration-space ensemble
admits the decomposition
\begin{equation}
\rho_{N,\lambda}(x)
=
\int da\,
\rho_{N,\lambda}(a)\rho_N(x\mid a).
\label{eq:mixture_conditionals}
\end{equation}
This is the standard representation of the
canonical ensemble as a probabilistic mixture of microcanonical ensembles
\cite{Touchette2015}.

Under extensivity assumptions, the tilted observable distribution
satisfies a large deviation  principle
\cite{Touchette2009,DemboZeitouni1998,Ellis1985},
\begin{equation}
\rho_{N,\lambda}(a)
\asymp
e^{-N I_\lambda(a)},
\end{equation}
with rate function
\begin{equation}
I_\lambda(a)
=
\lambda a-s(a)-\varphi(\lambda),
\label{eq:rate_function}
\end{equation}
where 
\begin{equation}
    s(a)=\lim_{N\to\infty}\frac{1}{N}\log\Omega_N(a),\quad \varphi(\lambda)=\lim_{N\to \infty} -\frac{1}{N}\log Z(\lambda)
\end{equation}
denote the entropy density and the free energy density, respectively.

If \(I_\lambda(a)\) has a unique minimizer \(a^\ast\), the distribution
\(\rho_{N,\lambda}(a)\) concentrates at \(a^\ast\):
\begin{equation}
\rho_{N,\lambda}(a)
\longrightarrow
\delta(a-a^\ast).
\label{eq:ldp_concentration}
\end{equation}
The convergence in Eq.~\eqref{eq:ldp_concentration} is written here at the level of the intensive macrostate distribution. If several minimizers retain comparable weight, the mixture need not collapse to a single conditional component; this is the probabilistic signature of coexistence and ensemble nonequivalence. More generally, ensemble equivalence can be formulated through several distinct notions of convergence, including convergence of probability measures, macrostates, and expectation values of observables. We refer to Ref.~\cite{Touchette2015} for a detailed discussion of these different levels of equivalence and the conditions under which they hold. 

Substitution of Eq.~\eqref{eq:ldp_concentration} into Eq.~\eqref{eq:mixture_conditionals} shows that the tilted ensemble becomes macroscopically dominated by the conditional ensemble at
\(a^\ast\). The soft bias imposed by exponential tilting is therefore
equivalent, at the level of the intensive macrostate, to a hard constraint on
its typical value. In the canonical--microcanonical setting, this is the
standard concentration mechanism underlying ensemble equivalence. Within the
present representation, it appears directly as the thermodynamic-limit
equivalence between soft reweighting and exact conditioning.

\section{Conclusion}

In this work, we have formulated reweighting as a unifying probabilistic
operation in statistical mechanics. Starting from a reference measure, a new ensemble is obtained by modifying the relative likelihood of configurations through a nonnegative weight.
Exponential tilting and conditioning are two elementary realizations of
this construction. The former incorporates information through expectation
constraints and redistributes probability continuously, whereas the latter
incorporates information through events or prescribed macrostates and restricts
the accessible support.

These transformations admit complementary variational interpretations. In one direction, the principle of minimum relative entropy determines the
updated ensemble, and therefore its reweighting factor, from the information
imposed on the system. Conversely, any prescribed nonnegative weight defines an
information potential and admits a variational representation. 
Strictly positive weights correspond to finite information
potentials, while vanishing weights appear as infinite penalties and therefore
implement hard support restrictions. Soft tilting and hard conditioning can thus be
viewed within a common variational structure.

The relative entropy also provides a common information-theoretic origin for the thermodynamic entropies. For a constant reference measure on the state space, minimum relative entropy under an
expectation constraint reduces to the maximization of Gibbs entropy. Under
conditioning, the relative entropy measures the surprisal of the constrained
macrostate. In the sharp-constraint limit, it separates into a universal
resolution-dependent contribution and a finite, macrostate-dependent part. The
latter coincides, up to an additive constant, with minus the Boltzmann entropy. Gibbs and Boltzmann entropy therefore arise from the same relative entropy framework.

The enlarged space representation makes the probabilistic relation between these
constructions explicit by promoting the observable to an explicit coordinate. The tilted ensemble in configuration space is obtained as a marginal of the observable tilted joint distribution, whereas the ensemble at fixed observable value is obtained as a conditional distribution of the same joint. In the thermodynamic specialization, the canonical ensemble appears as an exponentially reweighted marginal, whereas the microcanonical ensemble appears as a conditional ensemble.

In the thermodynamic limit, this representation acquires a
large deviation  interpretation. The tilted ensemble is, at every finite system
size, an exact probabilistic mixture of conditional ensembles. When the
distribution of the intensive observable concentrates around a unique typical
value, this mixture becomes dominated by a single conditional component.
Canonical--microcanonical equivalence can therefore be understood as the
macroscopic equivalence between soft reweighting and hard conditioning. 

The same probabilistic architecture extends beyond equilibrium configuration
space. On path space, static weights are replaced by trajectory functionals 
\cite{RoynetteYor2009, KellerBolhuis2024, Donati2017,  Singh2025}.
Conditioning generates bridge and transition-path ensembles and is naturally
related to committor functions and Doob-transformed dynamics, whereas
exponential trajectory tilting leads to Feynman--Kac ensembles and dynamical
large deviation  formalisms \cite{Doob1984,EVandenEijnden2010,TriplettLu2025,Dellago1998,Bolhuis2002}. More recently, change of measure on path space have been used to construct on
nonequilibrium theories the force--observable conjugate structures, generalizing the equilibrium thermodynamic formalism \cite{YangQian2020,Yang2026}.
This extension suggests that equilibrium ensembles, conditioned dynamics and
rare-event methods and non equilibrium thermodynamics are not separate constructions, but different applications of the same reweighting principle.



\section*{Acknowledgments}
I am deeply grateful to my PhD supervisor Christoph Dellago for the many stimulating discussions on statistical mechanics and stochastic processes. I also thank him for giving me the freedom to pursue this independent research. I would like to thank my colleagues Alessandro Coretti, Maximilian Negedly and Jonas Bojanovsky for their questions and fresh perspectives that repeatedly challenged my understanding and helped sharpen it. I also thank Ying-Jen Yang for constructive comments on the manuscript and for pointing out several relevant references that helped clarify the scope and presentation of this work.

During the preparation of this work, the author used OpenAI's ChatGPT, from GPT-4 to GPT-5.6, as an assistive tool for brainstorming, conceptual exploration, mathematical derivations, and writing and editing of the text. The author critically reviewed, verified, and independently validated all mathematical derivations, interpretations, and generated text. Full scientific responsibility for the accuracy and integrity of the manuscript remains entirely with the author.

This research was funded in part by the Austrian Science Fund (FWF) 10.55776/F81. For open access purposes, the author has applied a CC BY public copyright license to any author accepted manuscript version arising from this submission. I acknowledge the financial support by the Vienna Doctoral School in Physics (VDSP).

\bibliography{reweight_bibl}

\bibliographystyle{plain}

\appendix

\section{Thermodynamic Switches are Exponential Reweighting}
\label{app:switch}
Standard thermodynamic transformations can be expressed as reweighting operations. For instance:
\paragraph{Changing the intensive parameter.}
Given an exponential family $\rho_\lambda(x)$, a change $\lambda \to \lambda'$ can be written as ($
\Delta\lambda=\lambda'-\lambda$)
\begin{equation}
\rho_{\lambda'}(x)
=
\rho_{\lambda}(x)\,
\frac{
\exp\!\big(-\Delta \lambda A(x)\big)
}{
\left\langle
\exp\!\big(-\Delta\lambda A\big)
\right\rangle_{\lambda}
}
\end{equation}
Thus, changing temperature is exactly an exponential reweighting.
\paragraph{Controlling the extensive parameter.}
Similarly, controlling the observable with a field $A(x)\to A(x)+\mu B(x)$ yields
\begin{equation}
\rho_{\lambda,\mu}(x)
=
\rho_{\lambda}(x)\,
\frac{
\exp\!\big(-\lambda\mu B(x)\big)
}{
\left\langle
\exp\!\big(-\lambda\mu B\big)
\right\rangle_{\lambda}
}.
\end{equation}
Therefore, switching on a field is also an exponential reweighting.
These transformations are smooth changes of measure. In contrast, the passage from canonical to microcanonical ensembles involves exact conditioning and is therefore singular.

\section{Minimum Relative Entropy Calculations}
\label{app:min_rel}
This appendix derives the variational principles underlying the two fundamental reweighting operations discussed in the main text: exponential tilting and conditioning. In both cases, the transformed ensemble is obtained as the probability distribution that minimizes the relative entropy with respect to a reference measure while satisfying prescribed constraints.

\paragraph{Exponential Tilting as Minimum Relative Entropy}

Consider the variational problem  
\begin{equation}
\min_{\rho}
D(\rho\mid\rho_0)
\quad
\text{subject to}
\quad
\langle A\rangle_\rho=\bar A.
\end{equation}

Introducing Lagrange multipliers $\alpha$ and $\lambda$ for normalization and the expectation-value constraint, respectively, we extremize
\begin{equation}\begin{split}
\mathcal L[\rho]
&=
\int dx\,\rho(x)\log\frac{\rho(x)}{\rho_0(x)}\\&
+\lambda\left(\int dx\,\rho(x)A(x)-\bar A\right)\\&
-\alpha\left(\int dx\,\rho(x)-1\right). 
\end{split}
\end{equation}

Functional differentiation with respect to $\rho(x)$ gives
\begin{equation}
\rho(x)
=
\rho_0(x)\exp\!\bigl(- \lambda A(x)+\alpha-1\bigr).
\end{equation}
Normalization determines the constant $\alpha$ and yields the exponential family
\begin{equation}
\rho_\lambda(x)
=
\frac{e^{- \lambda A(x)}}{Z(\lambda)}\,\rho_0(x)
\end{equation}
with partition function
\begin{equation}
Z(\lambda)
=
\int dx\,\rho_0(x)e^{- \lambda A(x)}.
\end{equation}
The parameter $\lambda$ is fixed by the constraint
\begin{equation}
\langle A\rangle_\lambda
=
-\frac{\partial}{\partial\lambda}\log Z(\lambda)
=
\bar A.
\end{equation}

Therefore, exponential tilting is the least biased deformation of the reference ensemble compatible with a prescribed expectation value.

\paragraph{Conditioning as a minimum relative entropy problem}

We briefly show that conditioning is the minimum relative entropy solution
associated with a hard constraint. Let 
\begin{equation}
\bar A_\varepsilon=\{x:\bar A\le A(x)\le \bar A+\varepsilon\}\subset\Gamma
\end{equation} 
denote the admissible set of configurations after constraining on the finite resolution macrostate (see main text). Formally in the zero-width limit, this reduces to the level set \(A(x)=\bar A\).

Consider the variational problem
\begin{equation}
\min_{\rho} D(\rho\mid\rho_0),
\quad
\text{subject to}
\quad
\rho(x)=0 \ \text{for } x\notin \bar A_\varepsilon.
\end{equation}


Introducing a Lagrange multiplier \(\alpha\)
for normalization, we extremize
\begin{equation}\begin{split}
\mathcal{L}[\rho]
&=
\int_{\bar A _\varepsilon} dx\,\rho(x)\log\frac{\rho(x)}{\rho_0(x)}\\&
-
\alpha\left(\int_{\bar A _\varepsilon} dx\,\rho(x)-1\right).\end{split}
\end{equation}
Functional differentiation gives
\begin{equation}
\rho(x)=C\,\rho_0(x),
\qquad x\in {\bar A _\varepsilon}.
\end{equation}
Normalization fixes
\begin{equation}
C=\frac{1}{P_0(\bar A _\varepsilon)},
\qquad
P_0({\bar A _\varepsilon})=\int_{\bar A _\varepsilon} dx\,\rho_0(x).
\end{equation}
Thus the minimizer is 
\begin{equation}\label{eq:finite_cond}
\rho_{\bar A _\varepsilon}(x)
=
\frac{1_{\bar A _\varepsilon}(x)\rho_0(x)}
{P_0({\bar A _\varepsilon})}
=
\rho_0(x\mid {\bar A _\varepsilon}).
\end{equation}
Thus, conditioning is the least-biased distribution compatible with a support constraint. 
 
For every shell of finite width, the conditioned
ensemble remains absolutely continuous with respect to the reference
ensemble. In the sharp-shell limit, however, the measure becomes concentrated
on the level set $(A(x)=\bar A)$, which generally has zero measure under the
reference ensemble; the resulting conditional distribution becomes exactly the microcanonical conditioned distribution in Eq.~\ref{eq:conditioning} that is therefore singular
with respect to the ambient reference measure. Accordingly, Eq.~\eqref{eq:conditioning} should be understood as a distributional representation of a measure on the level set rather than as an ordinary
density on the original space.

\end{document}